\documentstyle[psfig]{l-aa}

\begin{document}

\thesaurus{ 06 (04.19.1, 08.14.1, 13.25.5) }

\title{ On the number of accreting and cooling isolated
neutron stars detectable with the ROSAT All-Sky Survey }

\author{R. Neuh\"auser \& J.E. Tr\"umper }

\offprints{R. Neuh\"auser, rne@mpe.mpg.de}

\institute{Max-Planck-Institut f\"ur extraterrestrische Physik, D-85740 Garching, Germany}

\date{Received date: 01 July 1998; accepted date: 13 Nov 1998 }

\maketitle

\markboth{ Neuh\"auser \& Tr\"umper:~~Isolated neutron stars detectable with ROSAT}{}

\begin{abstract}

We present limits to the $\log~N - \log~S$ curve for 
isolated neutron stars, both cooling and accreting neutron stars,
which are not active as radio pulsars, as observed with the ROSAT 
All-Sky Survey and compare it with theoretical expectations. 
So far, only one isolated neutron star is identified optically 
among ROSAT sources, namely RXJ185635-3754 (Walter \& Matthews 1997).
Three more promising candidates have been suggested.
In addition, several upper limit estimates are available on the 
space density of such neutron stars from different optical follow-up studies.
We show that the $\log~N - \log~S$ curve according to the current observations, 
including the identified neutron star, the three additional candidates, 
and the upper limits, 
lies between the theoretical expectations for middle-aged cooling 
neutron stars and old accreting neutron stars. At least one 
of the neutron star candidates found so far with ROSAT may be 
cooling instead of accreting. We suggest that the fact that more 
accreting isolated old neutron stars were expected (e.g., Madau \& Blaes 
1994) than observed is mostly due to the velocity distribution used 
in those calculations. More recent radio observations indicate 
that there are fewer slow neutron stars, ie., fewer accreting 
X-ray bright old neutron stars. At the X-ray bright end of the 
$\log~N - \log~S$ curve, however, the ROSAT observations agree 
well with the theoretical expectations.


\end{abstract}

\section{Introduction: Isolated neutron stars }

Both from the metallicity of the interstellar medium and the rate of supernovae 
(Narayan \& Ostriker 1990, van den Berg \& Tamman 1991, Tamman 1994),
it is expected that the Galaxy is populated by $10^{8}$ to $10^{9}$ neutron stars (NS);
of those, only $\sim 700$ have been discovered so far as radio pulsars 
or in binary systems (Taylor et al. 1993). Isolated old NS (IONS),
which are not active as radio pulsars (see Caraveo et al. 1996), 
may be visible as bright (up to $\sim 10 ^{32}~erg~s ^{-1}$) 
and soft ($k \cdot T \simeq 10$ to $\simeq 300~eV$) X-ray sources, 
for $T \simeq 10 ^{5}$ to $\simeq 10^{7}~K$.
They can emit X-rays either as old NS due to accretion of material 
from the interstellar medium (Ostriker et al. 1970) or 
as middle-aged NS due to emission from a cooling surface (eg., Becker 1995).

Depending on some model assumptions, a few hundred to several thousand
accreting IONS were expected to be detectable with the 
ROSAT All-Sky Survey (RASS) in the $0.1$ to $2.4~keV$ range 
(Treves \& Colpi 1991, henceforth TC91; Colpi et al. 1993; 
Blaes \& Madau 1993, henceforth BM93; Madau \& Blaes 1994, henceforth MB94). 
The fact that only one to four IONS (see Sect. 2 for details) have been 
found among RASS sources so far seems to be in contradiction with the expectations.
Eg., Livio et al. (1998) argue that this is due to a magnetic field
evolution different from what is assumed so far. However, as already 
pointed out by Neuh\"auser et al. (1997) and Haberl et al. (1997),
only the X-ray brightest IONS can be identified as such 
with currently available optical telescopes 
(see Sect. 3 for details).

The theoretical expectations by TC91 and MB94 are partly based on the 
distribution of the space velocity $v$ of NS at birth. 
Because Bondi-Hoyle accretion (Bondi \& Hoyle 1944) scales with $v ^{-3}$, 
only the slowest NS will be bright enough to be detectable in the RASS. 
Both the TC91 and MB94 calculations are based on the NS birth velocity 
distribution suggested by Narayan \& Ostriker (1990), but more recent 
studies indicate that there are significantly fewer NS with low velocities
(Lyne \& Lorimer 1994, Manning et al. 1996, Hansen \& Phinney 1998a, 1998b,
Cordes \& Chernoff 1998), so that there would consequently also
be fewer IONS detectable with the RASS (see Sect. 5 for details).

In this paper, we first summarize the observational facts on the number 
and space density of accreting IONS found in the RASS by compiling all the
data known for individual IONS candidates (Sect. 2) and listing
upper limits on the IONS space density obtained from follow-up 
identification programs (Sect. 3). In Sect. 4 we present several estimates 
for the local space density of NS and derive the $\log~N - \log~S$ curves 
for IONS observable with the RASS for both cooling and accreting NS. 
Finally, in Sect. 5, we compare these expectations with actual
observations and discuss, whether the apparent discrepancy may be due to the 
velocity distribution of NS being different from the one assumed in TC91 and MB94.

\section { The isolated neutron stars found so far }

The only candidate IONS from {\em Einstein Observatory} follow-up programs, 
namely MS0317.7-6647 (Stocke et al. 1995), shows a hard energy tail in its 
ASCA spectrum and may hence be a low-mass X-ray binary located in NGC 1313.
With the ROSAT PSPC, the whole sky has been observed down to a limiting flux 
of $\sim 10^{-13}~erg~s ^{-1}$. Voges et al. (1996) have published the RASS Bright 
Source Catalog called 1RXS with 18811 sources down to a limiting count rate 
of $0.05~cts~s ^{-1}$. For details on ROSAT and its prime instruments, 
the Positional Sensitive Proportional Counter (PSPC) 
and the High Resolution Imager (HRI), see Tr\"umper (1982), 
Pfeffermann et al. (1988), and David et al. (1996), respectively.

So far, only one IONS has been identified optically among RASS sources,
namely RXJ185635-3754 \footnote{also called RXJ1856.5$-$3754, 1RXS~J185635.1$-$375433, 
1RXP~J185635$-$3754.7, 1WGA~J1856.5$-$3754, 1ES1853$-$379, 1ES1853$-$37.9, 1RXH~J185635.3$-$375432} 
(Walter et al. 1996, Neuh\"auser et al. 1997, Walter \& Matthews 1997,
Neuh\"auser et al. 1998), a bright ($(3.67 \pm 0.15)~cts~s ^{-1}$ with the PSPC)
and soft ($k \cdot T \simeq 57~eV$ blackbody spectrum) previously unidentified RASS source.
It was first detected by the {\em Einstein Observatory} slew survey, then with RASS 
as well as PSPC and HRI pointed observations; all the different flux measurements
are consistent with constant X-ray emission, the flux being 
$(1.46 \pm 0.04) \cdot 10^{-11}~erg~s ^{-1} cm ^{-2}$. It has also been detected with the EUVE
(Lampton et al. 1997). Walter \& Matthews (1997) have identified the optical counterpart 
with the HST at $F_{606} = 25.8$ mag ($\simeq V$) and $F_{300} = 24.5$ mag ($\simeq U$),
ie., faint ($f_{X}/f_{V} \simeq 6 \cdot 10^{5}$) and blue as expected for IONS.
The object has also been detected from the ground with the ESO-3.5m-NTT at
$V = 25.7$ mag (Neuh\"auser et al. 1998). 

The object is located near the R CrA dark cloud area, most certainly 
inside or foreground to this cloud (see Fig. 2 in Neuh\"auser et al. 1997)
at a distance of $\le 130~pc$ (Marraco \& Rydgren 1981).
An upper limit to the radius can be obtained using the Greenstein \& Hartke (1983) 
model with $k \cdot T = 62~eV$ at the pole and $k \cdot T = 0~eV$ at the equator,
which yields the emitting area ($\le 2500~km ^{2}$) and $R_{\infty} \le 14~km$.
The small distance is consistent with the low absorbing column density 
found in a fit to the PSPC spectrum, namely $A_{V} = 0.1$ mag 
or $N_{H} = 1.4 \cdot 10^{20} cm ^{-2}$ (Walter et al. 1996).
The optical magnitudes are inconsistent with the 
non-magnetic H or He atmospheres predicted by Pavlov et al. (1996), 
and are also brighter than their pure black-body spectrum by $\sim 1$ mag, 
and fainter than their magnetic H atmospheres by $\sim 2$ mag. 
Although these inconsistencies might indicate an ion cyclotron emission feature, 
we do not favor this explanation, because the EUVE, $V$, $F_{300}$, and $F_{606}$ 
fluxes all follow a Planckian, which is somewhat more absorbed than 
by just $A_{V} = 0.1$ mag, which was obtained from the PSPC spectrum alone. 
Because this NS cannot be associated with any supernova remnant and 
because it does not show pulsation, it may well be an IONS accreting 
from the interstellar medium.

\begin{table*}

\begin{tabular}{lccccl}
\multicolumn{6}{c}{\bf Table 1. Upper limits on space density of RASS-detected IONS } \\ \hline
Projects and areas & area & number  & space density of  & count rate            & reference \\
            & $[deg^{2}]$ & of IONS & IONS per sr       & limit $[cts~s ^{-1}]$ &  \\ \hline 
Hamburg Quasar Survey     & 8480 & $\le 205$ & $\le 79$           & 0.05 & Bade et al. (1998) \\
Gal. Plane Survey Cygnus  & 64.5 & $\le 8~(\le 36)$ & $\le 407~~(1833)$  & 0.02 (0.012) & Motch et al. (1997a,b) \\
Northern areas I, IVac (Va)&216 (37)&$\le 8~(\le 10)$&$\le 122~~(888)$ & 0.03 (0.01) & Zickgraf et al. (1997) \\
High-latitude MBM clouds  & 64.5 & $\le 2$   & $\le 102$          & 0.012 & Danner et al. (1998a) \\
Dame et al. dark clouds & 1600   & $\le 1$   & $\le 2.1$          & 0.05  & Danner et al. (1998b) \\
Very bright soft sources & 30000 & $\le 3$   & $\le 0.33$         & 0.5 & Thomas et al. (1998) \\ \hline
\end{tabular}

\end{table*}

RXJ0720.4-3125 \footnote{also called 1RXS J072025.1$-$312554} is another NS
among previously unidentified RASS sources. It was first discovered in the ROSAT 
Galactic Plane Survey (Motch et al. 1991) and is similar to RXJ185635-3754.
Its PSPC spectrum with $(1.69 \pm 0.07)~cts~s ^{-1}$ shows a 
$k \cdot T \simeq 79~eV$ blackbody spectrum with low ($N_{H} \sim 10^{20} cm ^{-2}$)
absorption (Haberl et al. 1997). Its emission is modulated by a $8.39~sec$ pulse period.
Recently, Motch \& Haberl (1998) presented two optical candidate counterparts 
to this RASS source at $B \simeq 26.1$ (star X1) and $26.5$ mag (star X2), detected 
with the ESO-3.5m-NTT. Star X1 has also been detected by Kulkarni \& van Kerkwijk (1998),
their star X, with $B = (26.6 \pm 0.2)$ and $R = (26.9 \pm 0.3$) mag with the Keck II.
Hence, with the X-ray flux being $1.8 \cdot 10^{-11}~erg~s ^{-1} cm ^{-2}$,
the flux ratio $f_{X}/f_{opt}~\ge~2 \cdot 10^{5}$ excludes everything else
than an IONS. 

Another candidate IONS has been found in the galactic plane, 
namely RXJ0806.4-4123 \footnote{also called 1RXS J080623.0$-$412233}
(Haberl et al. 1998). 
The PSPC count rate of this source is
$(0.38 \pm 0.03)~cts~s ^{-1}$. The blackbody spectrum has 
$k \cdot T = (78 \pm 7)~eV$ for an absorbing column density of
$N_{H} = (2.5 \pm 0.9) \cdot 10^{20} cm ^{-2}$.
No optical counterpart has been found so far down to $B \sim 24$ mag 
(Haberl et al. 1998), so that $f_{X}/f_{opt}~\ge~2500$ may already exclude 
everything but an IONS as counterpart.

The RASS source RXJ1308.6+2127 
\footnote{also called 1RXS J130848.6$+$212708, RBS1223}
found during the ROSAT Bright Survey (RBS) by Schwope et al. (1999), 
with $f_{X}/f_{opt}~\ge~2 \cdot 10^{5}$ can be nothing else than an IONS.
This source is also  soft and, with $(0.287 \pm 0.024)~cts~s ^{-1}$ observed 
with the PSPC, it is the faintest IONS candidate found so far. No optical counterpart 
has been found down to $B \simeq 26$ mag inside the $1.5$ arc sec error circle 
around its HRI position (Schwope et al. 1999).

\section { Observed upper limits on the number of IONS }

There are several programs under way aiming at the identification of
RASS sources (compiled in Table 1). 
They all provide some constraints on the number density
of IONS: The number of unidentified sources can be taken as upper limit 
to the number of IONS. However, selection effects may bias these searches 
against finding IONS. For example, a bright star or AGN near the RASS source 
position may be incorrectly assumed to be the correct counterpart.
Hence, we have to add the number of possible mis-identifications to 
the number of unidentified sources to obtain the upper limit to the 
number of IONS. 
Following Bade et al. (1998) and Motch et al. (1997a,b), we allow $2~\%$ of possibly
wrong identifications for the remainder of the surveys, as listed below.
This is a conservative estimate, because the probability for random 
positional coincidence of an X-ray source with, eg. a CV, AGN, hot white dwarf 
or else, is $\le 10^{-3}$ at a PSPC count rate of $\ge 0.02~cts~s ^{-1}$.
This is extremly conservative, because neither will
all the mis-identificatified X-ray sources be IONS
nor are all the as yet unidentified sources really IONS.

The follow-up identification programs compiled in Table 1 are as follows:
\begin{itemize}
\item Bade et al. (1998), in the Hamburg Quasar Survey, search for 
optical identifications to RASS sources with $|b| \ge 20 ^{\circ}$. 

\item Motch et al. (1997a,b) have identified all but a few sources 
in their $64.5$ deg$^{2}$ Cygnus area. 

\item Zickgraf et al. (1997) have carried out an identification 
program for several northern high-latitude fields. 
In their areas I ($144$ deg$^{2}$) and IVac ($72$ deg$^{2}$), 
they could identify all but a few sources.

\item Danner (1998a) searched for IONS in all translucent clouds listed in 
Magnani et al. (1985). 

\item Danner (1998b) also searched for IONS in the dark clouds listed
by Dame et al. (1987), a total of $\sim 1600$ deg $^{2}$. He investigated 
and identified all the $16$ soft and bright RASS sources in or near these 
clouds, but found only one IONS candidate, namely RXJ185635-3754, 
which was already presented by Walter et al. (1996). 

\item For the ROSAT Bright Survey (Hasinger et al. 1997),
an identification program for all the $2012$ bright RASS sources 
with a count rate $\ge 0.2~cts~s ^{-1}$ in the $\sim 20000$ deg$^{2}$ 
area at $|b| \ge 30 ^{\circ}$ excluding the Virgo cluster and the 
Magellanic clouds, no reliable numbers on unidentified or possibly
mis-identified sources are available, yet. Within this program, 
however, Schwope et al. (1999) identified RXJ1308.6+2127 as IONS 
candidate as mentioned above. 

\item Thomas et al. (1998) identified all but one 
of the bright ($\ge 0.5~cts~s ^{-1}$),
soft (more counts in the $0.1-0.4~keV$ band than in the $0.5-2.1~keV$ band) 
RASS sources at $|b| > 20 ^{\circ}$, but did not find any IONS. 
\end{itemize}

Most of the expected NS have small X-ray fluxes and are faint optical sources.  
On the assumption that the optical magnitude of the only
reliably optically identified IONS (RXJ185635-3754) is typical for these
objects, an IONS near the RASS flux limit has $V \simeq 31$ mag.
However, Geminga shows optical emission brighter than expected
from a blackbody extrapolation of its X- and $\gamma$-ray spectrum, 
possibly an ion cyclotron line (Mignani et al. 1998). Hence, radio-quiet,
isolated (middle-aged and old) NS may actually be up to $\sim 2$ mag brighter
in the optical than expected from pure blackbodies. However, the claimed 
cyclotron emission feature on Geminga relies on only one data point,
and neither RXJ185635-3754 nor the three IONS candidates (not yet detected
optically) do show any indication for such an emission feature. This 
argues against optical emission features to be abundant in IONS.

The ESO-3.5m-NTT can achieve a limiting magnitude of $V = 27.4$ mag 
in an one hour exposure (at new moon, with $0.5$ arc sec seeing, 
for a $3~\sigma$ detection). Hence, with a ten hour exposure (a full 
long night), one can detect objects as faint as $V \simeq 29$ mag. 
With the Keck 10-m telescopes and the HST, the best currently available 
telescopes, one can achieve a limiting magnitude of $V \simeq 30$ mag 
(with similar parameters as above).
Using the ESO-8.2m-VLT telescopes UT1 and UT2, both simultaneously
with FORS~1 and FORS~2, one will be able to detect objects as faint as
$V \simeq 31$ mag when exposing for as long as a full winter night.
This is exactly the optical magnitude expected for an IONS
with X-ray count rate at the RASS flux limit.

\section { The $\log~N - \log~S$ curve of isolated NS }

We shall first compare estimates for the local space 
density of NS, and will then derive the expected $\log~N - \log~S$ 
curves of RASS detectable old accreting and middle-aged cooling NS. 
Then we compare these expectations with actual ROSAT observations, 
both detected IONS (and candidates) as well as upper limits on
their space density.

\subsection { The local space density of neutron stars }

One can estimate the number of IONS to be detectable with the RASS 
due to accretion by investigating Bondi-Hoyle mass accretion $\dot M$ 
onto an IONS travelling through the ambient interstellar material, 
which is assumed to be ionized. 

TC91 estimate the number density of IONS by a Monte-Carlo simulation
to be $\sim 1.5 \cdot 10^{6}~kpc ^{-2}$ (projected onto the galactic plane). 
Assuming Bondi-Hoyle accretion, TC91 derive the $\log~N-\log~S$ curve 
for RASS detectable IONS (see their Fig. 1), shown also in our Fig. 1.
Because the probability to discover such an accreting IONS is larger in 
regions with dense interstellar material, Colpi et al. (1993) list the 
numbers of IONS individually for a number of near-by galactic dark clouds.

BM93 derived the number density of NS to be 
$\sim 3 \cdot 10^{4}~kpc ^{-2}$, 
if there are a total of $10^{8}$ NS in the Galaxy.
These are NS with ages up to the age of the Galaxy, $\sim 10^{10}$ yrs.

To check whether the assumptions made for those calculations are correct, eg., 
regarding the local NS space density (or the total number of NS in the Galaxy),
we can compare those numbers with actual observations,
both radio and X-ray observations:
Lyne et al. (1998) estimated the number of unabsorbed NS, observable as 
radio pulsars, to be $(156 \pm 31)~kpc ^{-2}$ (corrected for beaming).
A NS is observable as radio pulsar up to an age of $\sim 10^{7}$ yrs.
Hence, extrapolating from the density of up to $10^{7}$ yr old NS
to the density of up to $10^{10}$ yr old NS, the NS density 
from radio observations is $(1.6 \pm 0.3) \cdot 10^{5}~kpc ^{-2}$.

The density of NS can also be estimated from NS found as 
X-ray point sources in supernova remnants. There are four 
such objects within $r \simeq 3$ kpc, namely Puppis A, G$296.5+0$, 
PKS $1209-52$, and G$117.7+0.6$ (see Caraveo et al. 1996).
Allowing for some obscuration by interstellar absorption, the real
number may be somewhat higher, say four to ten NS in supernova remnants
within $\sim 3$ kpc. Hence, their density is $\sim 0.17$ to $0.35~kpc ^{2}$.
These objects are up to $\sim 2 \cdot 10^{4}$ yrs old.
Hence, extrapolating to up to $10^{10}$ yr old NS, the NS density 
from X-ray observations is  $\sim 0.8$ to $1.5 \cdot 10^{5}~kpc ^{-2}$.

The numbers obtained by Lyne et al. (1998) from radio observations
and by us from X-ray observations are consistent with each other,
but the numbers obtained by TC91 and BM93 are somewhat different,
probably due to unrealistic assumptions made by TC91 and BM93.
We shall use $\sim 1.6 \cdot 10^{5}~kpc ^{-2}$ for the
remainder of this paper.

\begin{figure*}
\vbox{\psfig{figure=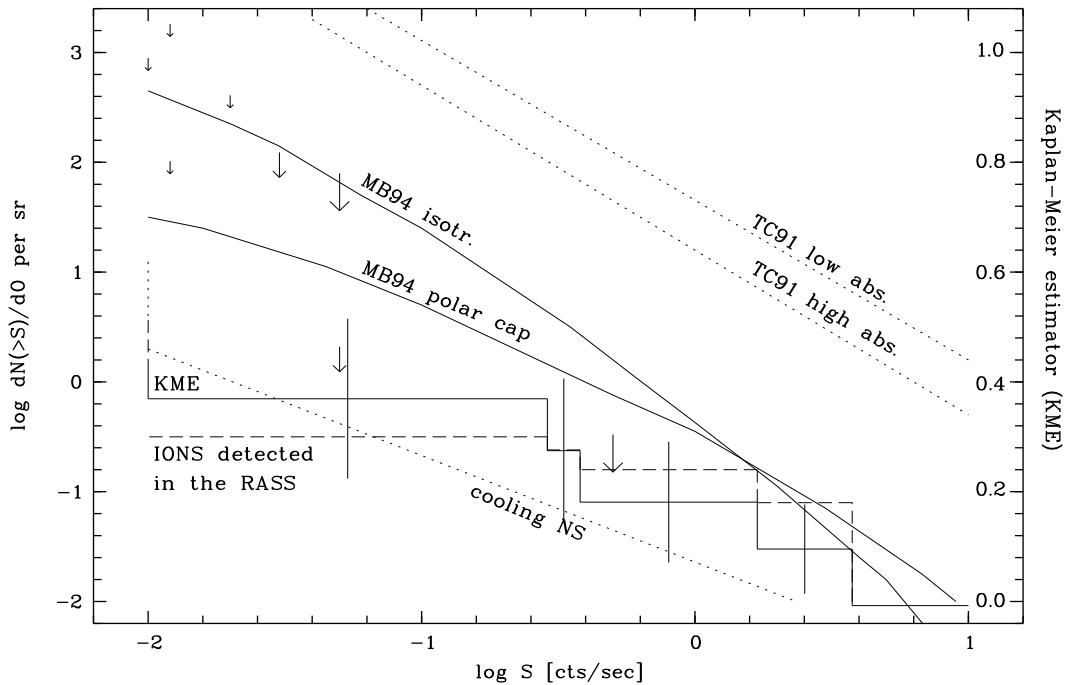,angle=-90,width=15cm,height=10cm}}
\caption{ Isolated NS in the RASS: Expectations and observations.
We plot the number density of (accreting or cooling) NS with ROSAT PSPC 
counts rate above $S$ (ie. $\log~(dN(>S)/d\Omega~sr ^{-1})$) versus the 
count rate $\log~S$, ie. $\log~N-\log~S$ curves.
For accreting IONS, we show the theoretical expectations by TC91 for both 
low and high absorption (upper dotted lines), and by MB94 for both 
polar cap and isotropic accretion (full lines).
Our expectation for cooling NS is also shown (lower dotted line). 
Then, we also plot the $\log~N-\log~S$ curve as observed for the four 
(candidate) IONS (broken line, four bins).
Also shown are upper limits from follow-up programs in limited sky areas; 
the length of the arrows are proportional to the sky area investigated (three bins); 
these upper limits are not hard limits, as they are obtained in limited areas. 
Then, we show the Kaplan-Meier estimator (KME) for the 
NS $\log~N-\log~S$ curve calculated from the four detections 
and the upper limits (full line, four bins with error bars) } 
\end{figure*}

\subsection { Accreting neutron stars detectable in the RASS }

Using the local space density of NS obtained in BM93,
and assuming $0.095~cm ^{-3}$ as density of the local interstellar material
within $100~pc$ around the Sun, ie. inside the local bubble, and $1~cm ^{-3}$
outside, MB94 obtain the number of IONS detectable along the 
line of sight above a ROSAT PSPC count rate $S$ to be
\begin{displaymath}
\frac{dN}{d \Omega} (> S)~=~n~\int_{0}^{\infty }~d \dot M~\int_{0}^{d(\dot M)}~dr~r^{2}~f(\dot M,r)
\end{displaymath}
with $d(\dot M)$ as maximum source distance and the IONS accretion rate distribution $f(\dot M,r)$.
MB94 plot the cumulative source counts expected for the RASS in their Fig. 2b,
also shown in out Fig. 1.

\subsection { Cooling neutron stars detectable in the RASS }

The standard NS cooling curve following the F-P model 
(Friedman \& Pandharipande 1981, Umeda et al. 1993) is consistent
with the observational data (see Becker \& Tr\"umper 1997), 
so that the X-ray luminosity $L_{X}$ of a cooling NS drops fast 
at $\sim 10^{6}$ yrs, ie. at $L_{X} = 10^{32}~erg~s ^{-1}$.
Therefore, in order to restrict the calculation to young, up to 
$10^{6}$ yrs old NS, we have to scale the space density given above,
assuming that the birth rate of NS has been constant in time.
Hence, the local space density of up to $10^{6}$ yr old,
ie., X-ray bright NS, is $\sim 16~kpc ^{-2}$ 
(or $\sim 7~kpc ^{-3}$ for a scale height of 450 pc).

For an object with a blackbody spectrum, the PSPC instrument detects 
X-ray photons most efficiently for spectra peaking at $k \cdot T \simeq 50~eV$. 
At this energy \footnote{ According to the Technical Appendix of the 
ROSAT Call for Proposals, the counts to energy conversion factor is
$\simeq 3.3 \cdot 10^{11}~cts~cm ^{2}~erg ^{-1}$ for $N_{H} = 10 ^{18}~cm ^{-2}$.}, 
the RASS detection limit of $\sim 0.01~cts~s ^{-1}$ 
corresponds to an X-ray flux of $\sim 3 \cdot 10 ^{-13}~erg~cm ^{-2}~s ^{-1}$,
if absorption is negligible. Hence, one can detect an unabsorbed NS 
with the RASS up to a distance of $\sim 5.2~kpc$.

Now, if we consider low absorption by $0.07$ H atoms $cm ^{-3}$ (Paresce 1984),
again to be conservative (in order not to underestimate the number of cooling NS),
one can detect a NS with $L_{X} = 10^{32}~erg~s ^{-1}$ down to the RASS limit
only up to a distance of $\sim 1.6~kpc$.
Hence, in a cylinder with $\sim 1.6~kpc$ radius and a scale-height of $450$ pc,
there are $\sim 25$ isolated cooling NS to be expected in the RASS 
down to the flux limit of $0.01~cts~s^{-1}$.
This corresponds to cooling NS number density of $\log~(dN/d\Omega~sr ^{-1})= 0.3$.
We display the corresponding $\log~N-\log~S$ curve in Fig. 1.

\section { Discussion and conclusion }

In Fig. 1, we compare the expectations by MB94 and TC91 for accreting NS 
and - as estimated above - for cooling NS with all the observations available 
so far, both the lower limits due to the four identified IONS 
(candidates) and the upper limits as compiled above. 
Taking into account the upper limits and the identified NS candidates,
ie., using the Kaplan-Meier estimator (computed with ASURV, Feigelson \& Nelson 1985,
Isobe et al. 1986), we get approximately the same $\log~N-\log~S$ curve
as for the four IONS (candidates) alone.

At or above the PSPC count rate of the faintest IONS candidate known 
so far ($0.287~cts~s^{-1}$), one would expect 
about one cooling isolated NS, and between one and four are found so far.
For those four, however, it is not known, yet, whether they are cooling,
middle-aged NS or accreting IONS.

The upper limits to the number density of IONS detected by RASS at the
fainter end, towards the RASS detection limit, do not rule out any of 
the models, just because these upper limits are obtained in limited areas,
ie., are not hard limits.

At the bright end of the $\log~N-\log~S$ curve, the lower limits derived 
from the four IONS candidates are consistent with the data because all bright 
RASS sources are already identified. The observations indicate that there are a few 
more IONS than expected from the $\log~N-\log~S$ curve of cooling middle-aged NS. 
At the X-ray bright end, the observed $\log~N-\log~S$ curve agrees well 
with the expectation by MB94 for accreting old NS, but the TC91 
expectations are far off. Also, the upper limits to the NS space density
obtained by ROSAT follow-up studies are consistent with the MB94
expectations, but inconsistent with TC91.

Motch et al. (1997a) also compared the $\log~N-\log~S$ curve of the unidentified
sources in their $64.5$ deg$^{2}$ area in Cygnus with the TC91 and MB94 expectations 
of the number density of IONS and concluded that the TC91 model seems to be excluded 
from the observational data, but the MB94 model may be consistent with the data.

Obviously, the large number of accreting NS expected by TC91 and MB94 
are not found in the data.
As discussed above (Sect. 3), the vast majority of these objects
would be faint in the optical and, hence, difficult to
identify optically with current technology.

The number of IONS detectable in the RASS depends crucialy on the velocity
distribution of NS, in particular at the low-velocity end, because Bondi-Hoyle
accretion scales with $v ^{-3}$. Both TC91 and BM93 assumed 
the best-fit velocity distribution obtained by Narayan \& Ostriker (1990), 
which shows a peak in their non-Maxwellian distribution at $\sim 40~km~s^{-1}$.
MB94 then assumed diffuse heating of the NS population,
so that the peak in their heated distribution was shifted towards 
higher velocities, namely to almost $100~km~s^{-1}$.
More recently, Lyne \& Lorimer (1994) and Lorimer et al. (1997),
using many newly published proper motions and a revised distance model, 
have published a new velocity distribution, which is different 
from the one by Narayan \& Ostriker (1990) in two important regards:
The mean velocity is larger in the new (Maxwellian) distribution,
namely $480~km~s^{-1}$, and the number of low-velocity NS is smaller.
However, Hartman (1997) argues that the initial NS velocity distribution
has more low-velocity NS than the distribution proposed by Lyne \& 
Lorimer (1994), if one takes into account errors in the proper motions.
Manning et al. (1996), Hansen \& Phinney (1998a, 1998b), and Cordes \& Chernoff
(1998) also come to the conclusion that there are significantly fewer
low velocity NS than assumed earlier.

Repeating the TC91 or MB94 calculations with the new velocity 
distribution would, however, be beyond the scope of this paper.
With less NS at low velocity, one would obviously expect less 
accreting IONS detectable with the RASS.

In any case, the optical follow-up observations are so far limited 
to the brightest objects. Within the errors of the currently 
observed IONS density and the upper limits presented, 
the RASS observations are in agreement with theoretical expectations.

\acknowledgements{
We would like to thank Frank Haberl, Werner Becker, G\"unther Hasinger, Axel Schwope, 
Norbert Bade, Christian Motch, H.-C. Thomas, and Fred Walter for usefull discussion. 
We are also grateful to our referee, Duncan Lorimer, for several 
useful comments, which improved this paper. ROSAT is supported 
by the German government (BMBF/DLR) and the Max-Planck-Society.  }

{}

\end{document}